\documentclass[aip, reprint, physrev]{revtex4-2}

\usepackage{graphicx}
\usepackage{dcolumn}
\usepackage{bm}
\usepackage[utf8]{inputenc}
\usepackage[T1]{fontenc}
\usepackage{amsmath}
\usepackage{mathptmx}
\usepackage{etoolbox}
\usepackage{accents}
\usepackage{colortbl}
\usepackage{enumitem}
\usepackage{comment}
\usepackage{booktabs}
\usepackage{array,multirow}
\usepackage[table,xcdraw]{xcolor}
\usepackage[version=3]{mhchem} 
\usepackage{orcidlink}

\newcommand{\kcalmol}{\mbox{kcal$\cdot$mol$^{-1}$}}
\newcommand{\kJmol}{\mbox{kJ$\cdot$mol$^{-1}$}}

\makeatletter
\def\@email#1#2{
 \endgroup
 \patchcmd{\titleblock@produce}
  {\frontmatter@RRAPformat}
  {\frontmatter@RRAPformat{\produce@RRAP{*#1\href{mailto:#2}{#2}}}\frontmatter@RRAPformat}
  {}{}
}
\makeatother
\begin{document}


\title[FNO quintuples]{FNO-CCSDTQ(5)$_\Lambda$ as an economical alternative for connected quintuple excitations contributions in coupled cluster thermochemistry}

\author{Gregory H. Jones*\orcidlink{0000-0003-3275-1661}}
    \email{ghjones@mit.edu}
\thanks{Equally contributing authors}

\affiliation{Quantum Theory Project, Department of Chemistry,
University of Florida, Gainesville, FL 32611, USA}
\affiliation{Present address: Department of Chemistry, Massachusetts Institute of Technology, Cambridge, MA 02139, USA}

\author{Aditya Barman\orcidlink{0009-0003-3863-2564}} 
\thanks{Equally contributing authors}
\affiliation{Department of Molecular Chemistry and Materials Science, Weizmann Institute of Science, 7610001 Re\d{h}ovot, Israel}

\author{Margarita Shepelenko\orcidlink{0000-0003-4707-1650}} 
\affiliation{Department of Molecular Chemistry and Materials Science, Weizmann Institute of Science, 7610001 Re\d{h}ovot, Israel}

\author{Jan M. L. Martin*\orcidlink{0000-0002-0005-5074}} 
\thanks{On sabbatical at the Quantum Theory Project}
    \email{gershom@weizmann.ac.il}    
    \affiliation{Department of Molecular Chemistry and Materials Science, Weizmann Institute of Science, 7610001 Re\d{h}ovot, Israel}

 \date{CPLETT-26-969 revision \today}

\begin{abstract}
Contributions from connected quintuple excitations in coupled cluster theory can reach the 0.5 \kcalmol\ range, important enough to matter in accurate computational thermochemistry, yet the very steep $\propto N^{12}$ CPU time scaling impedes routine evaluation. We show that for the differential contribution of quintuples, convergence of a frozen natural orbital (FNO) expansion with respect to the NO cutoff is rapid enough to make FNO-CCSDTQ(5)$_\Lambda$ with cutoffs of 0.0025 or 0.001 viable alternatives. A naive extrapolation to zero cutoff from \{0.005,0.0025\} works surprisingly well as a low-cost option. Interestingly, FNO convergence is definitely slower for second-row than for first-row compounds.
\end{abstract}

\maketitle

\section{Introduction}\label{sec:intro}

Total atomization energies (TAEs) are the thermochemical cognates of heats of formation. Computing TAEs at the sub-\kJmol\ accuracy level required for such projects as Active Thermochemical Tables (ATcT)\cite{Ruscic2021ActiveEffects,Ruscic2013a,Ruscic2022} requires accounting for post-CCSD(T) correlation effects. (For reaction energies that involve less drastic chemical transformations, judicious use of hierarchies of (homo/iso)desmotic reactions --- see, e.g., Refs.\citenum{wheeler2009,RaghavachariCBH2011} --- may obviate the need of most higher-order corrections.)

Valence post-CCSD(T) corrections are routinely evaluated in such protocols as the HEAT (high-accuracy extrapolated \emph{ab initio} thermochemistry) approach of the late lamented Stanton and coworkers\cite{HEAT,HEAT2,HEAT3,HEAT4} and W4 (Weizmann-4) of Karton et al.\cite{jmlm200,jmlm205}, as well as its explicitly correlated\cite{jmlm269} and $\Lambda$-coupled cluster\cite{lambdastanton1,lambdastanton2,lambdabartlett1,lambdabartlett2} variants.\cite{jmlm330} 
Thanks to the NCC (New Coupled Cluster\cite{NCC1,NCC2}) module, built on the TBLIS tensor contraction library,\cite{TBLIS} in the CFOUR\cite{CFOUR} electronic structure program system, CCSDT(Q),\cite{mrcc8} CCSDT(Q)$_\Lambda$,\cite{KallayGauss2005,KallayGauss2008} and full CCSDTQ\cite{CCSDTQ} calculations can now be carried out with unprecedented efficiency. 

However, for still higher connected excitations, the only practically available implementation remains the fully general coupled cluster code in MRCC.\cite{MRCC}




Significant contributions of connected quintuple excitations for some molecular properties were first reported by Musia{\l} et al.\cite{Musial2001} (see also Refs.\citenum{Thogersen2004,Ruden2004}). Karton et al.\cite{jmlm200} in 2006 reported that the total atomization energy (TAE, i.e., the sum of all bond energies) of ozone has a connected quintuples component of 0.3--0.5 \kcalmol. Both the HEAT protocol\cite{HEAT,HEAT2,HEAT3,HEAT4} and the W4 (Weizmann-4) approach\cite{jmlm200,jmlm205} include connected quintuples contributions.

The extremely steep CPU time scaling $\propto O^5V^7$ for CCSDTQ5 (where $O$ and $V$ represent numbers of occupied and virtual orbitals, respectively), and $\propto O^5V^6$ for CCSDTQ(5) and CCSDTQ(5)$_\Lambda$, makes the evaluation of $T_5$ contributions a very costly affair. In the course of attempting to assess the importance of quintuples for a well-known multireference organic chemistry problem (\emph{vide infra}) we ran into a computational scaling wall. As a field expedient, we attempted to carry out the same calculation in a reduced set of virtuals --- which ultimately led us to the present study. 

The idea of frozen natural orbitals (FNO) ultimately can be traced back to the iterative natural orbitals approach of Bender and Davidson.\cite{BenderDavidson1969} Following the realization that iteratively refining natural orbitals has very poor convergence properties in practice, Meyer\cite{Meyer1973} proposed a single-shot set of NOs followed by configuration interaction (CI) in that reduced basis set; Sosa et al.\cite{Bartlett1989} considered the same for many-body methods.

FNO-CC in the modern sense of the word was first proposed by Taube and Bartlett\cite{Taube2005,Taube2008} for ground-state coupled cluster and by Krylov\cite{Krylov2010,Krylov2020} for excited-state EOM-CC (equations-of-motion coupled cluster). The method was also implemented\cite{RolikKallay2011} in the MRCC\cite{MRCC} program system.



\section{Computational Details}\label{sec:methods}


All electronic structure calculations in this work were carried out on the CHEMFARM HPC system at the Weizmann Institute of Science. The FNO-CCSDTQ(5)$_\Lambda$ calculations have been carried out using the arbitrary-order coupled cluster code\cite{KallaySurjan2001,KallayGauss2005,KallayGauss2008} in a development version of MRCC\cite{MRCC} kindly provided by Prof. Mih\'aly K\'allay.

The only basis sets considered in the present work are the Dunning correlation consistent\cite{Dunning1989,Woon1993} polarized double, triple,  quadruple, and quintuple-zeta basis sets (cc-pVDZ, cc-pVTZ,  cc-pVQZ, and cc-pV5Z), as well as the cc-pVDZ and cc-pVTZ basis sets with the highest angular momentum removed, which we denote cc-pVDZ(p,s) and cc-pVTZ(d,p), respectively.

Reference geometries for the W4-08\cite{jmlm215} 
thermochemical benchmark were taken verbatim from the ESI of the respective references. However, in order that we could reuse the CCSDTQ(5)$_\Lambda$/cc-pVDZ data from the W5preview paper\cite{jmlm340}, the species in Table~\ref{tab:fno_vdz_virtuals} were evaluated at the CCSD(T)/cc-pwCVQZ optimized geometries from the ESI of the said paper. For subsets of W4-08, CCSDTQ(5)$_\Lambda$/cc-pVTZ and cc-pVTZ(d,p) calculations were carried out at the same geometries in this work.

Some CCSDTQ/cc-pVDZ and CCSDT(Q)$_\Lambda$/cc-pVTZ calculations on the benzyne isomers were carried out using a development version of the CFOUR\cite{CFOUR} electronic structure program system. 





In principle, one could evaluate (5)$_\Lambda$ contributions to the total atomization energy; however, with FNO this prompts the question as to size consistency, since the sum of the numbers of FNOs for the separated atoms often does not match the number of molecular FNOs retained. However, for quintuple excitation contributions this is a nonissue for any compound in the \{H,B,C,Al,Si\} system as the atomic quintuples contribution will be identically zero. For the remaining elements, as seen in Table~\ref{tab:atoms_postccsdtq}, quintuples contributions to the atomic energies are quite negligible, with the possible exception of chlorine atom, and even the latter would drown in remaining possible error sources (not least of which are anharmonic zero-point corrections). We have hence elected to carry out the comparison entirely in terms of the molecular correlation energy increments.

\begin{table}[t]
\centering
\caption{Post-CCSDT(Q)$_\Lambda$ corrections (\kcalmol) for the atoms at full cc-pVTZ basis set and different FNO cutoffs.}
\label{tab:atoms_postccsdtq}
\small
\begin{tabular}{lccccccc}
\hline
&(5)$_\Lambda$ & \multicolumn{6}{c}{FNO(cutoff)-(5)$_\Lambda$}\\
\cline{3-8} \\
\multicolumn{2}{c}{}& $10^{-4}$
& $2.5\times10^{-4}$
& $5\times10^{-4}$
& $10^{-3}$
& $2.5\times10^{-3}$
& $5\times10^{-3}$\\

\hline

N	&	0.000	&	0.000	&	0.000	&	0.000	&	0.000	&	0.000	&	0.000\\
O	&	0.001	&	0.001	&	0.001	&	0.001	&	0.001	&	0.001	&	0.000\\
F	&	0.001	&	0.002	&	0.002	&	0.002	&	0.002	&	0.001	&	0.000\\
P	&	0.001	&	0.001	&	0.001	&	0.001	&	0.001	&	0.000	&	0.000\\
S	&	0.002&	0.002$_5$	&	0.002$_5$	&	0.003	&	0.002	&	0.002	&	0.001\\
Cl	&	0.004	&	0.004	&	0.004	&	0.004	&	0.003	&	0.003	&	0.002\\
\hline
\end{tabular}
For B, C, Al, Si corrections are 0 by definition if cores are frozen.
\end{table}

\section{Results and discussion}

\subsection{W4-08  thermochemical benchmark}

\begin{table*}[t]
\centering
\caption{Convergence for selected W4-08 species of quintuples correction to TAE (\kcalmol) as a function of basis set or FNO threshold; percentages of full virtual space for different FNO thresholds; statistics over the whole dataset. The cc-pVDZ basis set was used. 
CCSD(T)/cc-pwCVQZ reference geometries and full virtual space results were taken from the ESI of Ref.\citenum{jmlm340}.
}\label{tab:fno_vdz_virtuals}
\scriptsize
\setlength{\tabcolsep}{4pt}

\begin{tabular}{lccccccccc}
\hline

&
\multicolumn{1}{c}{Full virtual space}
&
\multicolumn{4}{c}{FNO virtual space retained}
&
\multicolumn{4}{c}{Percentage of VDZ virtual space (\%)} \\

\cline{2-2}
\cline{3-6}
\cline{7-10}

System
& VDZ{\@}pwCVQZ$^b$
& FNO($5\times10^{-4}$)
& FNO($10^{-3}$)
& FNO($2.5\times10^{-3}$)
& FNO($5\times10^{-3}$)
& $5\times10^{-4}$
& $10^{-3}$
& $2.5\times10^{-3}$
& $5\times10^{-3}$ \\

\hline
\ce{FOOF}$^c$ & 0.309 & 0.291 & 0.281 & 0.205 & 0.137 & 94.2 & 90.4 & 55.8 & 44.2 \\
\ce{P4} & 0.211 & 0.197 & 0.198 & 0.158 & 0.099 & 88.5 & 86.5 & 65.4 & 48.1 \\
\ce{SO2} & 0.115 & 0.117 & 0.111 & 0.095 & 0.092 & 94.9 & 82.1 & 61.5 & 46.2 \\
\ce{SO3} & 0.122 & 0.122 & 0.121 & 0.100 & 0.086 & 90.4 & 78.8 & 57.7 & 44.2 \\
\ce{CO2} & 0.059 & 0.068 & 0.059 & 0.056 & 0.048 & 84.6 & 71.8 & 53.8 & 38.5 \\
\ce{CS2} & 0.098 & 0.099 & 0.094 & 0.090 & 0.059 & 92.3 & 87.2 & 71.8 & 48.7 \\
\ce{OCS} & 0.071 & 0.074 & 0.071 & 0.065 & 0.067 & 92.3 & 79.5 & 59.0 & 43.6 \\
\ce{S2O} & 0.143 & 0.142 & 0.133 & 0.117 & 0.086 & 94.9 & 89.7 & 71.8 & 51.3 \\
\ce{O3} & 0.430 & 0.421 & 0.386 & 0.285 & 0.221 & 94.9 & 87.2 & 56.4 & 43.6 \\
\ce{S3} & 0.169 & 0.165 & 0.154 & 0.141 & 0.085 & 94.9 & 92.3 & 79.5 & 59.0 \\
\ce{S4} ($C_{2v}$) & 0.429 & 0.414 & 0.414 & 0.358 & 0.209 & 94.2 & 94.2 & 80.8 & 55.8 \\
\ce{ClCN} & 0.127 & 0.119 & 0.104 & 0.074 & 0.028 & 87.2 & 74.4 & 66.7 & 43.6 \\
\ce{HNO} & 0.083 & 0.083 & 0.072 & 0.050 & 0.023 & 87.1 & 74.2 & 54.8 & 35.5 \\
\ce{NO2} & 0.181 & 0.181 & 0.165 & 0.134 & 0.105 & 94.9 & 79.5 & 56.4 & 41.0 \\
\ce{N2O} & 0.196 & 0.180 & 0.168 & 0.135 & 0.074 & 87.2 & 76.9 & 59.0 & 38.5 \\
\ce{B2} & 0.055 & 0.053 & 0.053 & 0.036 & 0.022 & 80.8 & 65.4 & 50.0 & 30.8 \\
\ce{F2O} & 0.098 & 0.102 & 0.098 & 0.065 & 0.034 & 94.9 & 92.3 & 53.8 & 43.6 \\
\ce{Cl2O} & 0.080 & 0.081 & 0.081 & 0.057 & 0.036 & 94.9 & 94.9 & 76.9 & 61.5 \\
\ce{BN} & 0.177 & 0.183 & 0.185 & 0.290 & 0.316 & 80.8 & 61.5 & 53.8 & 30.8 \\
\ce{HCN} & 0.098 & 0.089 & 0.082 & 0.069 & 0.009 & 74.2 & 64.5 & 54.8 & 32.3 \\
\ce{CCH} & 0.067 & 0.062 & 0.057 & 0.016 & 0.007 & 71.0 & 64.5 & 45.2 & 29.0 \\
\ce{C2H2} & 0.075 & 0.069 & 0.063 & 0.048 & 0.005 & 63.9 & 58.3 & 47.2 & 27.8 \\
\ce{C2} & 0.339 & 0.357 & 0.348 & 0.188 & 0.166 & 80.8 & 73.1 & 50.0 & 30.8 \\
\ce{N2} & 0.111 & 0.110 & 0.083 & 0.083 & 0.011 & 88.5 & 65.4 & 65.4 & 34.6 \\
\ce{P2} & 0.115 & 0.114 & 0.106 & 0.104 & 0.066 & 96.2 & 88.5 & 73.1 & 50.0 \\
\ce{CO} & 0.040 & 0.042 & 0.039 & 0.035 & 0.069 & 84.6 & 69.2 & 57.7 & 34.6 \\
\ce{CN} & 0.118 & 0.118 & 0.089 & 0.051 & 0.020 & 84.6 & 65.4 & 57.7 & 34.6 \\
\ce{NO} & 0.084 & 0.082 & 0.073 & 0.064 & 0.040 & 92.3 & 76.9 & 61.5 & 38.5 \\
\ce{O2} & 0.096 & 0.097 & 0.090 & 0.053 & 0.018 & 96.2 & 88.5 & 57.7 & 42.3 \\
\ce{S2} & 0.052 & 0.053 & 0.053 & 0.041 & 0.018 & 96.2 & 96.2 & 80.8 & 57.7 \\
\ce{SO} & 0.063 & 0.064 & 0.062 & 0.050 & 0.030 & 96.2 & 92.3 & 73.1 & 50.0 \\
\ce{F2} & 0.039 & 0.043 & 0.043 & 0.027 & 0.009 & 96.2 & 96.2 & 57.7 & 46.2 \\
\ce{Cl2} & 0.019 & 0.022 & 0.022 & 0.020 & 0.006 & 96.2 & 96.2 & 92.3 & 65.4 \\
\hline
Geom.mean&&&&&&80.4	&70.8	&54.7	&38.1\\
Max.&&&&&&97.4	&96.2	&92.3	&65.4\\
Min.&&&&&&53.6	&42.4	&37.5	&24.2\\
\hline
RMSD & REF$^a$ & 0.007 & 0.013 & 0.037 & 0.065\\
\hline
\end{tabular}

(a) RMS value is 0.113 \kcalmol; (b) values taken from ESI of Ref.\citenum{jmlm340}; (c) Evaluated at CCSD(T)/cc-pV(Q+d)Z geometries from ESI of Ref.\citenum{jmlm273}.
\end{table*}

Complete results are available in the ESI. Table~\ref{tab:fno_vdz_virtuals} presents FNO-CCSDTQ(5)$_\Lambda$ data with the cc-pVDZ basis set for a representative subsample, compared with untruncated CCSDTQ(5)$_\Lambda$/cc-pVDZ. The final columns also display what percentage of the full number of virtuals is included at a given FNO threshold.

\begin{figure}[t]

\centering

\includegraphics[width=\columnwidth]{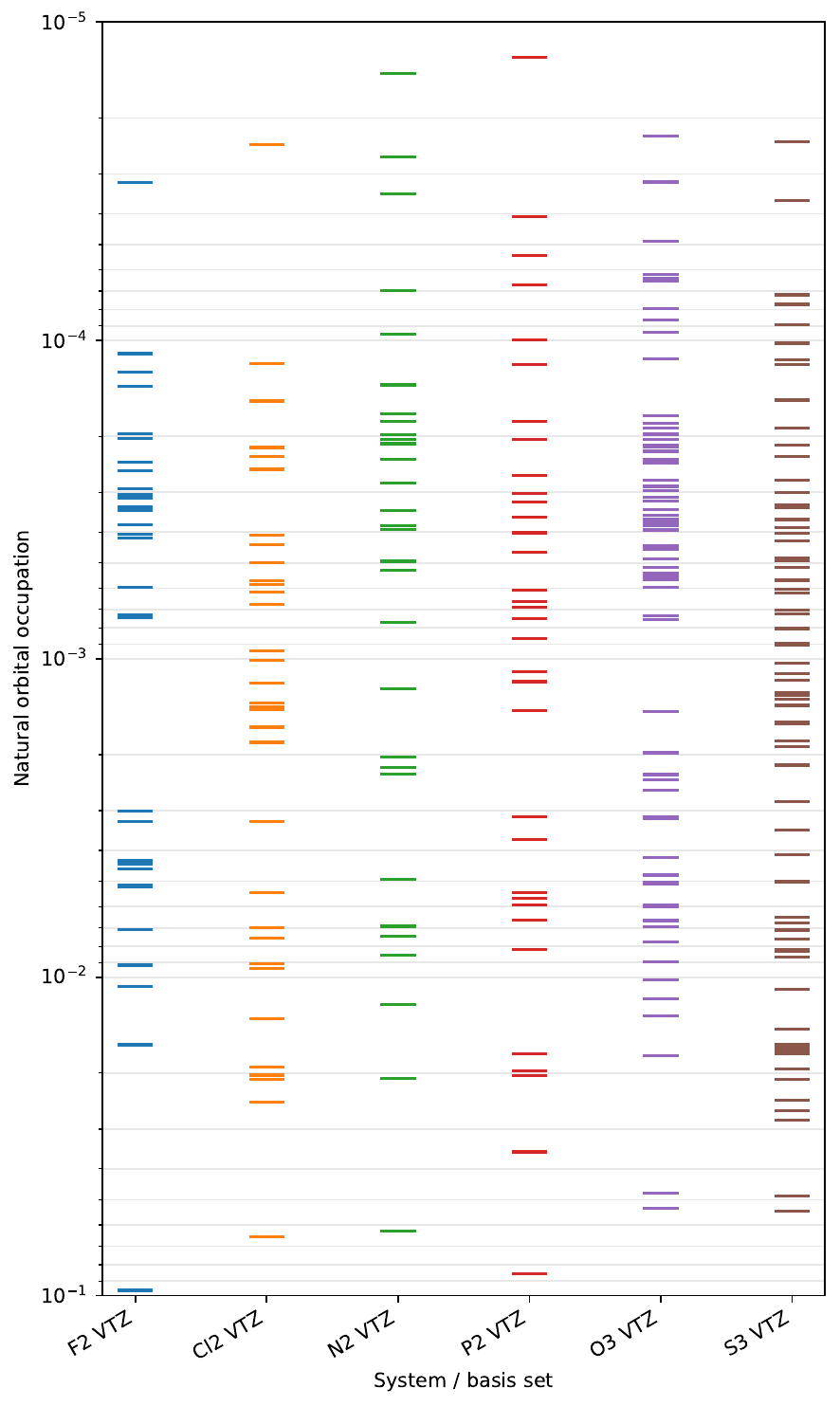}

\caption{
Natural orbital occupations of some isovalent species pairs for the VTZ basis set. Horizontal markers denote individual occupations.}

\label{fig:noons}

\end{figure}

A cutoff of 0.01 proved to yield too erratic results, with just a single virtual included in some cases, hence we abandoned it. FNO(0.005) includes a geometric mean of 37\% of the virtual orbital space, but with a very wide spread --- from just 25\% for \ce{BH3} and \ce{AlH3} to a whopping 65\% for \ce{Cl2}. Saturated hydrides aside, the second-row species clock in at markedly higher percentages than their first-row counterparts. This is probably best illustrated by Figure~\ref{fig:noons}.

\begin{figure}[t]

\centering

\includegraphics[width=\columnwidth]{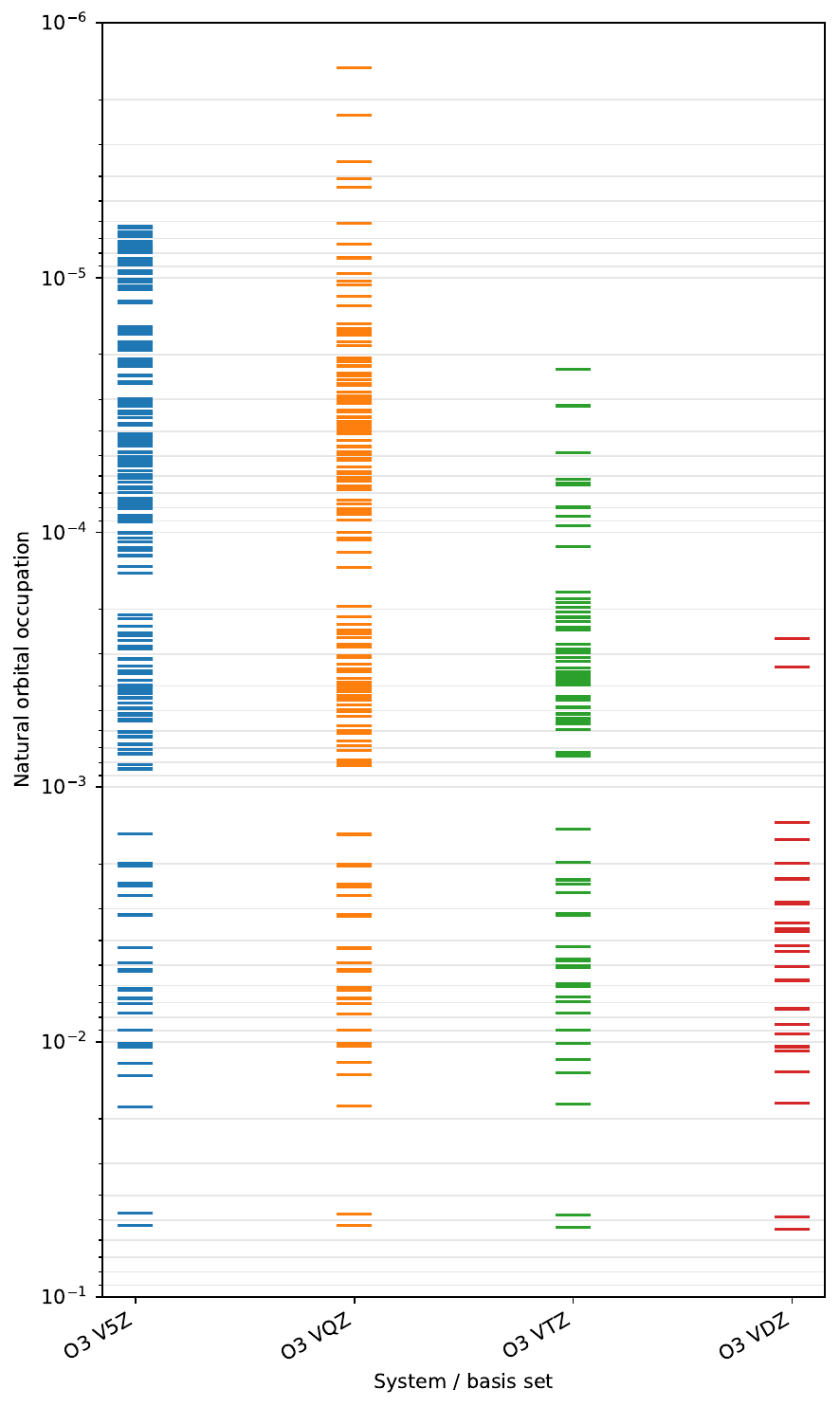}

\caption{
Natural orbital occupations of ozone with different basis sets. Horizontal markers denote individual occupations.}

\label{fig:o3_noons}

\end{figure}

FNO(0.005) only recovers about half the quintuples contribution for cases like \ce{P4}, \ce{O3}, \ce{S4}. FNO(0.0025) performs markedly better for such species, but seems to add little for \ce{C2}, \ce{SO2}, \ce{SO3}. A clear performance leap occurs for FNO(0.001), with the FNO result often being very close to the full-basis number. It should be noted that, while 71\% is the geometric mean percentage of the full orbital space which is included in the FNO calculations, the spread runs from 42\% for methane to 96\% for \ce{Cl2}, with the second-row species again tending to higher percentages than their first-row counterparts. We attempted pushing further to FNO(0.0005), but clearly this only avails for (mostly first-row) cases where 0.001 retained just 60-70\% of the full space. Comparison of NO histograms between cc-pVDZ and cc-pVTZ for the same species (see Figure~\ref{fig:o3_noons} for the example of ozone) reveals that the extra basis functions in cc-pVTZ make an entry at or just below FNO(0.001), so we conclude that cutoffs below 0.001 are not meaningful unless one also wants to upgrade the underlying basis set to cc-pVTZ.  (In Figure~\ref{fig:o3_noons}, a similar phenomenon is apparent for cc-pVQZ and cutoffs below 0.0001, as well as for cc-pV5Z and 0.00001, respectively.) Indeed, as seen in Table~\ref{tab:fno_vtz_virtuals}, already FNO(0.001) in a cc-pVTZ basis set can retain more FNOs than there would be virtuals available in a cc-pVDZ basis set.

Energy-wise, FNO(0.001) is actually a quite good approximation of the CCSDTQ(5)$_\Lambda$/cc-pVDZ results, with RMSD=0.013 \kcalmol\ out of an RMS quintuples contribution of 0.113 \kcalmol. 

So let us now switch gears to the cc-pVTZ basis set (Table~\ref{tab:fno_vtz_virtuals}). FNO(0.001) here achieves a comparable accuracy, RMSD=0.016 out of 0.099 \kcalmol. Tightening to FNO(0.0005) reduces the RMSD to 0.011, and further to 0.007 for FNO(0.00025). For the nonhydrogen diatomics, we pushed one step beyond to FNO(0.0001): only for \ce{C2} and \ce{BN} did this change results even just at the 0.02 \kcalmol\ level.

The largest quintuples contributions in our sample exceed 0.4 \kcalmol\ for ozone and tetrasulfur, and 0.3  \kcalmol\ for dicarbon. Several more species have contributions of 0.2  \kcalmol\, such as \ce{P4}, \ce{NO2}, \ce{N2O}, and the like. These are also the species for which FNO(0.0025) still falls a bit short, while this appears to be adequate for the species with smaller contributions.

Typically, the quintuples contribution to the TAE rises as the FNO threshold is tightened, then plateaus close to the full-basis values, with sometimes a mild oscillation remaining. An exception is the notoriously multireference BN diatomic, where convergence is actually from above --- coarse cutoffs lead to an \emph{overestimate} of its importance. 

A reviewer inquired about error estimates. One possibility would, of course, be to take twice the RMS error as an approximate 2-sigma error estimate. Another, more tailored to the specific system at hand, would be to consider the change between two successive FNO thresholds; this is somewhat akin philosophically to the focal point approach (FPA).\cite{schuurman2004} 

Can somewhat useful results be obtained through extrapolation from fairly inexpensive FNO(0.0025)/cc-pVDZ and FNO(0.005)/cc-pVDZ values? We experimented with a Petersson-style pair natural orbital extrapolation,\cite{NydenPetersson1981,Petersson1981a} to no avail. Then we attempted an empirical two-point extrapolation of the form E[0.0025]+$a\times$(E[0.0025]-E[0.005]): surprisingly, simply setting $a=1$ (in sort-of a naive Richardson extrapolation) turned out to yield the optimum RMSD=0.024 \kcalmol, which may prove useful for somewhat larger systems such as those discussed anon.


A reviewer requested timings. In response, Figure~\ref{fig:FNO_WallClock} present a double-log plot of wall clock times on 16 Intel Ice Lake cores for three representative molecules. As can be seen there, the effective CPU time scaling is fairly close to a power law in the number of FNOs retained, with effective exponents hovering around 6.

\begin{figure}[t]

\centering

\includegraphics[width=\columnwidth]{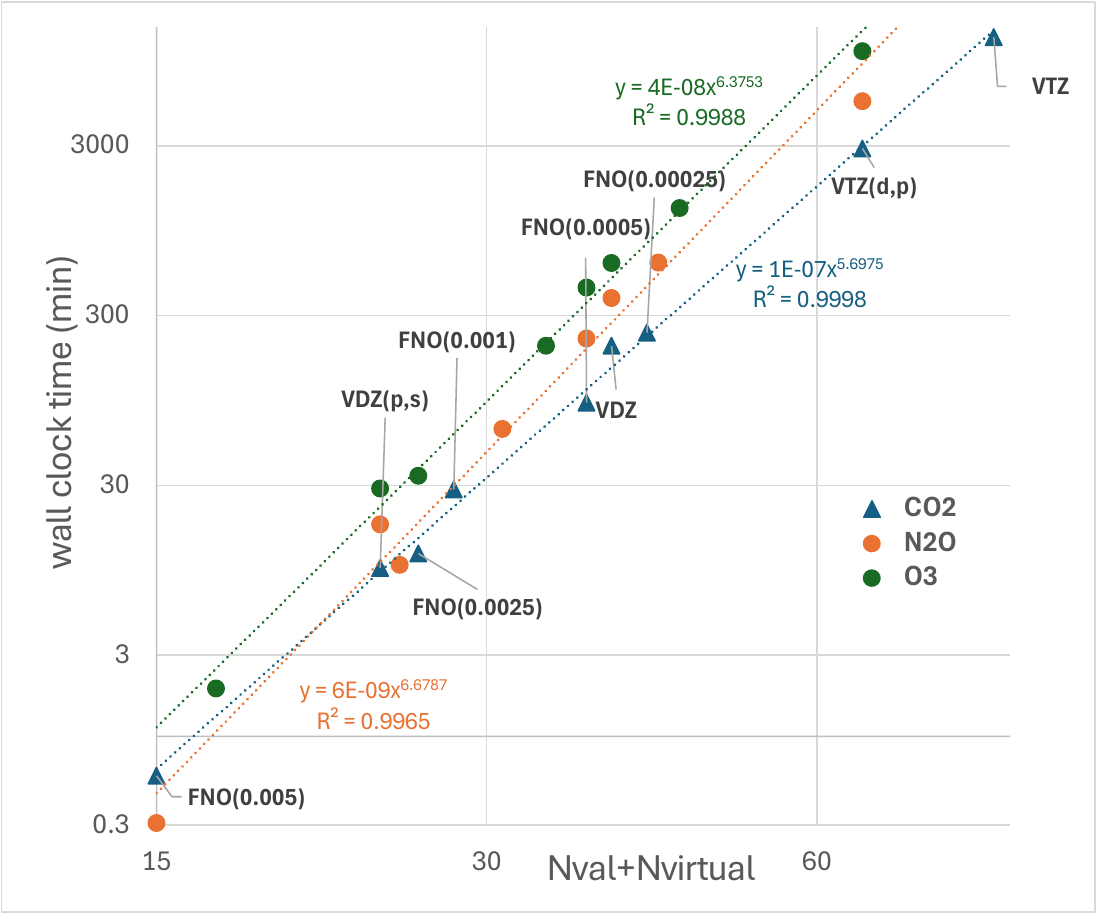}

\caption{CCSDTQ(5)$_\Lambda$ wall clock times (min) on 16 Intel Ice Lake cores for \ce{CO2}, \ce{N2O}, and \ce{O3} compared for different FNO(cutoff)/cc-pVTZ and different basis sets. Note that the three molecules may retain slightly different numbers of FNOs at a given cutoff.}

\label{fig:FNO_WallClock}

\end{figure}

\begin{table*}[t]
\centering
\caption{Convergence for selected W4-08 species of quintuples correction to TAE (\kcalmol) as a function of basis set or FNO threshold; percentages of full virtual space for different FNO thresholds; statistics over the whole dataset. The cc-pVTZ basis set was used. CCSD(T)/cc-pwCVQZ reference geometries and full virtual space results were taken from Ref.\citenum{jmlm340}.}
\label{tab:fno_vtz_virtuals}
\scriptsize
\setlength{\tabcolsep}{3pt}
\begin{tabular}{lcccc|cccccc|ccccccc}
\hline
&
\multicolumn{4}{c}{Full virtual space} &
\multicolumn{6}{c}{FNO Threshold} &
\multicolumn{7}{c}{\% cc-pVTZ Virtuals} \\
\cline{2-5} \cline{6-11} \cline{12-18}
System
& VTZ
& VTZ(d,p)
& VDZ
& VDZ(p,s)
& $10^{-4}$
& $2.5\times10^{-4}$
& $5\times10^{-4}$
& $10^{-3}$
& $2.5\times10^{-3}$
& $5\times10^{-3}$
& $10^{-4}$
& $2.5\times10^{-4}$
& $5\times10^{-4}$
& $10^{-3}$
& $2.5\times10^{-3}$
& $5\times10^{-3}$ 
& VDZ\\
\hline
\ce{FOOF} & --- & --- & 0.309 & 0.274 & --- & --- & \textcolor{red}{0.287} & 0.276 & 0.221 & 0.134 & --- & 52.6 & 42.2 & 40.5 & 30.2 & 19.8 & 44.8 \\
\ce{P4} & --- & --- & 0.211 & 0.129 & --- & --- & --- & 0.244 & 0.196 & 0.123 & --- & 63.8 & 50.9 & 40.5 & 29.3 & 21.6 & 44.8 \\
\ce{SO2} & --- & 0.135 & 0.115 & 0.150 & --- & 0.124 & 0.115 & 0.109 & 0.104 & 0.091 & --- & 55.2 & 46.0 & 39.1 & 33.3 & 21.8 & 44.8 \\
\ce{SO3} & --- & --- & 0.122 & 0.170 & --- & --- & --- & 0.115 & 0.113 & 0.084 & --- & 54.3 & 44.8 & 38.8 & 33.6 & 20.7 & 44.8 \\
\ce{CO2} & 0.062 & 0.060 & 0.059 & 0.061 & --- & 0.069 & 0.065 & 0.060 & 0.066 & 0.046 & --- & 48.3 & 42.5 & 32.2 & 29.9 & 17.2 & 44.8 \\
\ce{CS2} & 0.137 & 0.115 & 0.098 & 0.055 & --- & --- & 0.135 & 0.127 & 0.113 & 0.072 & --- & 62.1 & 51.7 & 39.1 & 32.2 & 21.8 & 44.8 \\
\ce{OCS} & --- & 0.077 & 0.071 & 0.055 & --- & --- & 0.087 & 0.083 & 0.082 & 0.071 & --- & 56.3 & 47.1 & 35.6 & 29.9 & 19.5 & 44.8 \\
\ce{S2O} & --- & --- & 0.143 & 0.153 & --- & \textcolor{red}{0.168} & 0.154 & 0.143 & 0.137 & 0.090 & --- & 64.4 & 51.7 & 40.2 & 35.6 & 23.0 & 44.8 \\
\ce{O3} & --- & 0.471 & 0.430 & 0.404 & --- & 0.441 & 0.421 & 0.386 & 0.305 & 0.217 & --- & 51.7 & 42.5 & 39.1 & 29.9 & 19.5 & 44.8 \\
\ce{S3} & --- & 0.216 & 0.169 & 0.121 & --- & \textcolor{red}{0.226} & 0.207 & 0.184 & 0.170 & 0.099 & --- & 73.6 & 58.6 & 43.7 & 36.8 & 26.4 & 44.8 \\
\ce{S4} ($C_{2V}$) & --- & --- & 0.429 & 0.286 & --- & --- & \textcolor{red}{0.520} & 0.469 & 0.414 & 0.250 & --- & 71.6 & 58.6 & 44.8 & 37.1 & 25.9 & 44.8 \\
\ce{ClCN} & 0.131 & 0.126 & 0.127 & 0.110 & --- & --- & 0.134 & 0.123 & 0.072 & 0.036 & --- & 57.5 & 47.1 & 37.9 & 29.9 & 20.7 & 44.8 \\
\ce{HNO} & 0.090 & 0.086 & 0.083 & 0.073 & --- & 0.091 & 0.090 & 0.078 & 0.054 & 0.023 & --- & 44.4 & 37.5 & 31.9 & 25.0 & 15.3 & 43.1 \\
\ce{NO2} & --- & 0.190 & 0.181 & 0.195 & --- & 0.183 & 0.177 & 0.167 & 0.144 & 0.103 & --- & 50.6 & 42.5 & 35.6 & 29.9 & 18.4 & 44.8 \\
\ce{N2O} & --- & 0.198 & 0.196 & 0.194 & --- & 0.197 & 0.197 & 0.176 & 0.141 & 0.072 & --- & 49.4 & 42.5 & 35.6 & 28.7 & 17.2 & 44.8 \\
\ce{B2} & 0.065 & 0.059 & 0.055 & 0.054 & 0.060 & 0.059 & 0.058 & 0.059 & 0.047 & 0.021 & 50.0 & 37.9 & 36.2 & 29.3 & 24.1 & 13.8 & 44.8 \\
\ce{F2O} & --- & --- & 0.098 & 0.089 & --- & 0.120 & 0.103 & 0.103 & 0.059 & 0.040 & --- & 50.6 & 42.5 & 42.5 & 28.7 & 20.7 & 44.8 \\
\ce{Cl2O} & --- & --- & 0.080 & 0.041 & --- & --- & 0.104 & 0.099 & 0.075 & 0.049 & --- & 69.0 & 58.6 & 42.5 & 36.8 & 27.6 & 44.8 \\
\ce{BN} & 0.218 & 0.189 & 0.177 & 0.190 & 0.199 & 0.220 & 0.217 & 0.204 & 0.287 & 0.312 & 62.1 & 41.4 & 36.2 & 27.6 & 24.1 & 13.8 & 44.8 \\
\ce{HCN} & 0.092 & 0.088 & 0.098 & 0.091 & --- & 0.096 & 0.096 & 0.093 & 0.069 & 0.009 & --- & 43.1 & 31.9 & 30.6 & 23.6 & 13.9 & 43.1 \\
\ce{CCH} & 0.063 & 0.059 & 0.067 & 0.058 & --- & 0.067 & 0.064 & 0.064 & 0.020 & 0.007 & --- & 43.1 & 30.6 & 30.6 & 20.8 & 12.5 & 43.1 \\
\ce{C2H2} & 0.065 & 0.061 & 0.075 & 0.063 & --- & 0.066 & 0.071 & 0.071 & 0.048 & 0.005 & --- & 40.7 & 29.1 & 26.7 & 19.8 & 11.6 & 41.9 \\
\ce{C2} & 0.351 & 0.327 & 0.339 & 0.270 & 0.341 & 0.363 & 0.360 & 0.351 & 0.187 & 0.164 & 65.5 & 43.1 & 36.2 & 32.8 & 22.4 & 13.8 & 44.8 \\
\ce{N2} & 0.109 & 0.104 & 0.111 & 0.112 & 0.107 & 0.111 & 0.118 & 0.112 & 0.087 & 0.012 & 74.1 & 48.3 & 39.7 & 37.9 & 29.3 & 15.5 & 44.8 \\
\ce{P2} & 0.152 & 0.127 & 0.115 & 0.091 & 0.145 & 0.139 & 0.127 & 0.113 & 0.112 & 0.072 & 81.0 & 63.8 & 50.0 & 39.7 & 36.2 & 22.4 & 44.8 \\
\ce{CO} & 0.041 & 0.039 & 0.040 & 0.046 & 0.044 & 0.051 & 0.044 & 0.047 & 0.041 & 0.068 & 67.2 & 44.8 & 39.7 & 34.5 & 29.3 & 15.5 & 44.8 \\
\ce{CN} & 0.118 & 0.110 & 0.118 & 0.060 & 0.079 & 0.085 & 0.084 & 0.066 & 0.066 & 0.033 & 65.5 & 48.3 & 37.9 & 29.3 & 29.3 & 15.5 & 44.8 \\
\ce{NO} & 0.087 & 0.084 & 0.084 & 0.087 & 0.087 & 0.091 & 0.088 & 0.086 & 0.069 & 0.051 & 81.0 & 50.0 & 41.4 & 37.9 & 31.0 & 19.0 & 44.8 \\
\ce{O2} & 0.110 & 0.106 & 0.096 & 0.085 & 0.109 & 0.108 & 0.106 & 0.106 & 0.067 & 0.042 & 87.9 & 50.0 & 43.1 & 43.1 & 32.8 & 22.4 & 44.8 \\
\ce{S2} & 0.078 & 0.068 & 0.052 & 0.025 & 0.076 & 0.074 & 0.070 & 0.066 & 0.061 & 0.034 & 91.4 & 74.1 & 58.6 & 43.1 & 39.7 & 29.3 & 44.8 \\
\ce{SO} & 0.079 & 0.077 & 0.063 & 0.050 & 0.078 & 0.074 & 0.068 & 0.064 & 0.061 & 0.036 & 89.7 & 58.6 & 51.7 & 41.4 & 37.9 & 24.1 & 44.8 \\
\ce{F2} & 0.071 & 0.069 & 0.039 & 0.031 & 0.070 & 0.066 & 0.048 & 0.048 & 0.033 & 0.006 & 91.4 & 51.7 & 43.1 & 43.1 & 31.0 & 20.7 & 44.8 \\
\ce{Cl2} & 0.038 & 0.036 & 0.019 & 0.004 & 0.037 & 0.036 & 0.035 & 0.032 & 0.030 & 0.013 & 94.8 & 82.8 & 63.8 & 43.1 & 41.4 & 29.3 & 44.8 \\
\hline
Geom.mean& & & & & & & & & & & ---& 47.2 & 38.0 & 31.5 & 26.1 & 16.9 & 43.1\\
Max.& & & & & & & & & & & --- & 82.8 & 63.8 & 44.8 & 41.4 & 29.3 & 44.8\\
Min.& & & & & & & & & & & --- & 29.6 & 21.1 & 16.5 & 15.3 & 9.4 & 38.8\\
\hline
RMSD & REF$^a$ & 0.008 & 0.014 & 0.026 & 0.006 & 0.007 & 0.011 & 0.016 & 0.037 & 0.057\\
\hline
\end{tabular}

(a) RMS value is 0.099 \kcalmol
\end{table*}

\begin{table}[h]
\caption{Convergence of sextuples corrections to TAE (\kcalmol) as a function of basis set or FNO threshold}
\label{tab:sextuples}
\centering
\begin{tabular}{lrrrrr}
\toprule
& \ce{C2} & \ce{BN} & \ce{P2} & \ce{N2O} & \ce{O3} \\
\multicolumn{6}{c}{E[$(6)_\Lambda$]=E[CCSDTQ5$(6)_\Lambda$]--E[CCSDTQ5]}\\
\midrule
FNO(0.005)/pVTZ & 0.062 & 0.020 & 0.006 & 0.013 & 0.025\\
FNO(0.0025)/pVTZ & 0.063 & 0.045 & 0.012 & 0.019 & 0.046\\
FNO(0.001)/pVTZ & 0.043 & 0.039 & 0.012 & 0.020 & 0.048\\
FNO(0.0005)/pVTZ & 0.046 & 0.040 & 0.013 \\
FNO(0.00025)/pVTZ & 0.048 & 0.039\ & 0.014\\
FNO(0.0001)/pVTZ & 0.052 & 0.035 & 0.015\\
cc-pVDZ & 0.052 & 0.032 & 0.012\\
cc-pVDZ(p,s)\cite{jmlm330} &0.054 & 0.033 & 0.021 &0.026&0.053\\
\midrule
\multicolumn{6}{c}{E[CCSDTQ5$(6)_\Lambda$]--E[CCSDTQ$(5)_\Lambda$]}\\
\midrule
FNO(0.005)/pVTZ &  0.059 &-0.029 & 0.020&0.009&0.029\\
FNO(0.0025)/pVTZ & 0.063 & 0.002 & 0.018&0.007&0.032 \\
FNO(0.001)/pVTZ & 0.029 & 0.003 & 0.016&0.003&0.021\\
FNO(0.0005)/pVTZ & 0.029 &-0.002 & 0.016\\
FNO(0.00025)/pVTZ & 0.031 &-0.005 & 0.017\\
FNO(0.0001)/pVTZ & 0.037 &-0.022 & 0.017\\
cc-pVDZ & 0.034 & -0.011 & 0.018\\
cc-pVDZ(p,s)\cite{jmlm330} & 0.021 & 0.004 & 0.028 & 0.007 & 0.041\\
\bottomrule
\end{tabular}

\end{table}

A reviewer wondered whether the FNO approach would also work for connected sextuples, E[$(6)_\Lambda$]=E[CCSDTQ5$(6)_\Lambda$]--E[CCSDTQ5]. Canonical results with the unpolarized cc-pVDZ(p,s) basis set can be found in the ESI of Ref.\cite{jmlm330}: only a handful of systems in W4-08 have nontrivial E[$(6)_\Lambda$] to begin with, and even those do not exceed 0.05 kcal/mol (for \ce{C2} and \ce{O3}). Furthermore, since we cannot take advantage here of a split between faster and slower codes the way we could for E[CCSDTQ$(5)_\Lambda$]--E[CCSDT$(Q)_\Lambda$], the more meaningful quantity is really E[CCSDTQ5$(6)_\Lambda$]--E[CCSDTQ$(5)_\Lambda$]. Some results for \ce{C2}, \ce{BN}, \ce{P2}, \ce{N2O}, and \ce{O3} are presented in Table~\ref{tab:sextuples}. If anything, it appears that convergence with respect to the FNO threshold is still faster than for quintuples; on the other hand, the values in question are so small, particularly for E[CCSDTQ5$(6)_\Lambda$]--E[CCSDTQ$(5)_\Lambda$], that their calculation is only desirable if all other contributions to the total atomization energy have been truly tied down. If however one is curious about the importance of sextuples in a species but cannot afford even a cc-pVDZ(p,s) calculation, FNO(0.001) or even FNO(0.0025) may be alternatives to consider.

\subsection{Proof of principle: the benzyne isomers}

\begin{table}[h]
\caption{Post-CCSD(T) total atomization energy contributions (\kcalmol) for benzyne isomers.}
\label{tab:benzyne_postccsdtq}
\centering
\begin{tabular}{lcccc}
\toprule
 & o-benzyne & m-benzyne & p-benzyne \\
\midrule
$\Delta$TAE[$T_3-$(T)]/cc-pVDZ             & -1.35  & -2.15  & -3.83  \\
$\Delta$TAE[$T_3-$(T)]/cc-pVTZ             & -2.75  & -3.60  & -5.16  \\
$\Delta$TAE[$T_3-$(T)]/cc-pV\{D,T\}Z             & -3.45  & -4.33  & -5.82  \\
$\Delta$TAE[(Q)]/cc-pVDZ                   & 2.66   & 3.78   & 7.18     \\
$\Delta$TAE[(Q)]/cc-pVTZ                   & 3.18   & 4.39   & 8.51     \\
$\Delta$TAE[(Q)]/cc-pV\{D,T\}Z             & 3.40   & 4.65   & 9.06     \\
$\Delta$TAE[(Q)$_\Lambda-$(Q)]/VDZ          & -0.06  & -0.22  & -1.65    \\
$\Delta$TAE[$T_4-$(Q)$_\Lambda$]/VDZ          & -0.38  & -0.59  & -1.04    \\
FNO(0.005)/(5)$_\Lambda$/VDZ & 0.088 & 0.175 & 0.283 \\
FNO(0.0025)/(5)$_\Lambda$/VDZ & 0.190 & 0.334 & 0.531 \\
(5)$_\Lambda$ extrap. \{0.005,0.0025\} & 0.29 & 0.49 & 0.78 \\
FNO(0.001)/(5)$_\Lambda$/VDZ & --- & --- & 0.701 \\
\bottomrule
\end{tabular}
$T_3-(T)$ and (Q) extrapolated from cc-pV\{D,T\}Z according to Table 5 in Karton\cite{Karton2020}
\end{table}

The isomers of benzyne (didehydrobenzene) are a classic multireference problem: for a detailed review of theoretical and experimental work on these species, see Vu et al.\cite{Lischka2024_benzyne}

\begin{table}[h]
\caption{Static correlation diagnostics (dimensionless) for benzyne isomers.}
\label{tab:benzyne_diag}
\centering
\begin{tabular}{lcccc}
\toprule
 & o-benzyne & m-benzyne & p-benzyne \\
\midrule
\%TAE[(T)] cc-pVTZ       & 2.65   & 3.02   & 4.17      \\
$\mathcal{T}_1$ CCSD/cc-pVTZ  & 0.0129 & 0.0155 & 0.0188 \\
$\mathcal{D}_1$ CCSD/cc-pVTZ  & 0.0390 & 0.0459 & 0.0632 \\
$\mathcal{D}_2$ CCSD/cc-pVTZ   & 0.2392 & 0.2672 & 0.4066 \\
$M_{\rm diag}$  CCSD/cc-pVTZ           & 0.113 & 0.137 & 0.291  \\
$I_\textrm{ND}^\textrm{max}$ CCSD/cc-pVTZ           & 0.0553 & 0.0670 & 0.1257 \\
$\overline{I_\textrm{ND}}$ CCSD/cc-pVTZ            & 0.0504 & 0.0516 & 0.0608\\
$\Delta I_\textrm{ND}^\textrm{max}$[(T)] &0.0185&0.0283&0.0775\\
$\Delta \overline{I_\textrm{ND}}$[(T)] cc-pVTZ& 0.0113 & 0.0129 & 0.0207 \\
$r_I$[(T)] cc-pVTZ  & 0.488 & 0.517 & 0.705 \\
$\Delta M_{\rm diag}$[(T)]& 0.042 & 0.067 & 0.270\\
DAD(CCSD)/cc-pVTZ  & 0.0061 & 0.0071 & 0.0083 \\
DAD(CCSDT)/cc-pVDZ & 0.0012 & 0.0015 & 0.0025  \\
max$|T_2|$ CCSD/cc-pVTZ  & 0.153 & 0.188 & 0.353  \\
max$|T_2|$ CCSDT/cc-pVTZ  & 0.194 & 0.249 & 0.552  \\
max$|T_2|$ CCSD/cc-pVDZ  & 0.169 & 0.210 & 0.385  \\
max$|T_2|$ CCSDT/cc-pVDZ  & 0.210 & 0.269 & 0.575  \\
max$|T_2|$ CCSDTQ/cc-pVDZ  & 0.215 & 0.279 & 0.601  \\
\bottomrule
\end{tabular}
\end{table}

All three species have singlet ground states. However,\cite{Lineberger1998Benzyne} while the singlet-triplet splitting in o-benzyne and m-benzyne is quite substantial, it drops to just a few \kcalmol\ for p-benzyne, which in organic chemistry classes is typically treated as a biradical.

For these calculations, we used the CCSD(T)/cc-pVQZ optimized geometries from Ref.\citenum{jmlm220} Results are presented in Table~\ref{tab:benzyne_postccsdtq}, while static correlation diagnostics can be found in Table~\ref{tab:benzyne_diag}.

The $\mathcal{T}_1$ and \%TAE[(T)] diagnostics\cite{Lee1989,jmlm200} for p-benzyne are actually deceptively low, just 0.019 and 4.2\%, respectively. These suggest some static correlation but nothing like the alarming largest doubles amplitude max|$T_2$|, which rises from 0.35 at the CCSD level to a startling 0.55 at the CCSDT level (with two triples amplitudes of 0.012), rising still further to 0.60 at the CCSDTQ level (the two largest triples amplitudes now being 0.014 and 0.013). No such drastic `jump' from CCSD to CCSDT is seen in o- and m-benzyne; since the newly proposed\cite{jmlm339} $\Delta \overline{I_\textrm{ND}}$[(T)] and $r_I$[(T)] are both essentially proxies for the change in the density upon introduction of connected triples, these diagnostics do give fair warning about the p-isomer's much more pronounced multireference character. Incidentally, so the doubles-based $\mathcal{D}_2$ diagnostic\cite{Nielsen1999a}, Matito's\cite{Ramos-Cordoba2016} $I_\mathrm{ND}^\mathrm{max}$ and especially $\Delta I_\mathrm{ND}^\mathrm{max}$[(T)] introduced in Ref.\citenum{jmlm339},  Truhlar's M diagnostic\cite{Tishchenko2008} from CCSD natural orbital occupations, and the change in M upon introducing (T). ($\mathcal{D}_2$ was evaluated using MOLPRO\cite{MOLPRO}.)

Full CCSDTQ/cc-pVDZ using CFOUR on 16 Intel Ice Lake cores took about 2.5 hours per iteration for the $D_{2h}$ p-benzyne, and about 4 hours per iteration for the lower-symmetry ($C_{2v}$) o-benzyne and m-benzyne isomers, compared about 10 hours per iteration for the p-benzyne triplet. (Geometries were optimized at the CCSD(T)/cc-pVQZ level using MOLPRO.)

The differences between CCSDTQ and CCSDT(Q)$_\Lambda$ are all negative, as expected\cite{jmlm330}, becoming more so from -0.38 for \emph{ortho} to -1.04 \kcalmol\ for \emph{para}. It is known from experience\cite{jmlm326,jmlm330}that a degree of error compensation exists between higher-order quadruples and connected quintuples; this is borne out here too. 

A full CCSDTQ(5)$_\Lambda$/cc-pVDZ(p,s) calculation proved elusive, but we were able to complete FNO(0.0025)/cc-VDZ for all three species. A simple linear extrapolation from FNO(0.005) and FNO(0.0025) to FNO(0) suggests quintuples contributions of \{0.3,0.5,0.8\} \kcalmol\ for \{\emph{ortho,meta,para}\}-benzyne: these cancel the lion's share of the $T_4-$(Q)$_\Lambda$ contribution but not all of it.

FNO(0.001) proved possible only for the \emph{para} isomer (thanks to its $D_{2h}$ computational symmetry). We obtained 0.701 \kcalmol, which is quite close to the value extrapolated at much smaller expense from the \{0.005,0.0025\} data points, 0.78 \kcalmol. 

It is noteworthy that across all three isomers --- as indeed seen more broadly for general thermochemistry\cite{jmlm340} and spectroscopic properties\cite{jmlm326} ---  connected quintuples cancel to a large extent with $T_4-(Q)_\Lambda$. This once more shows `the power of Lambda' (John F. Stanton, 1961-2025).\cite{jmlm326}

Overall, the sum of all the post-CCSD(T) correlation terms works out to: ortho -0.21, meta +0.01, para +1.34 \kcalmol. The corresponding totals if we neglected all post-CCSDT(Q)$_\Lambda$ terms  would have been -0.11, +0.10, +1.60 \kcalmol.

\section{Conclusions}

Connected quintuple excitations can be computationally more important than generally appreciated. For molecules with significant static correlation such as ozone, their importance in the total atomization energy can exceed 0.4 \kcalmol. Unfortunately, the steep CPU time scaling of CCSDTQ5 as $O^5V^7$, or of CCSDTQ(5)$_\Lambda$ as $O^5V^6$, soon puts this out of reach for all but the smallest molecules.

However, FNO-CCSDTQ(5)$_\Lambda$ turns out to be a viable approach to surmount this daunting scaling through truncation of a frozen natural orbital space. Since the contribution of connected quintuples to atomic total energies is either zero (for B, C, Al, Si) or negligible (N--Ne, P--Ar), this truncation will not cause (meaningful) deviations from size-consistency.

FNO(0.0025)/cc-pVDZ appears to be the lowest-cost alternative that provides a meaningful estimate. FNO(0.001)/cc-pVDZ or FNO(0.001)/cc-pVTZ will yield results quite close to the full orbital space limit. With cc-pVDZ and cc-pVTZ basis sets, cutoffs below 0.001 and 0.0001, respectively, are contraindicated as they enter NO occupation regions where basis functions from the next larger basis set are truly required.

A naive extrapolation to zero cutoff from \{0.005,0.0025\} works surprisingly well, and (as demonstrated for the p-benzyne example) may be an attractive low-cost option for not-so-small molecules.

Aside from saturated hydrides, our results clearly indicate that FNO(Threshold) calculations on second-row species require a significantly larger percentage of the full orbital space than for the isovalent first-row molecules. This is consistent with the often-stated observation (see, e.g., Ref.\citenum{jmlm340}) that basis set convergence is slower for second-row species than for first-row species.

We illustrate the utility of the FNO-CCSDTQ(5)$_\Lambda$ approach through the isomers of benzyne. Connected quintuple excitations are found to reach nearly 1 \kcalmol\ for the para isomer, and to largely, but not perfectly, cancel with higher-order connected quadruples contributions.

\subsection*{CRediT authorship contribution statement}
GHJ: conceptualization (equal), methodology (supporting), data interpretation (equal);   writing - review and editing (equal);
ADB: investigation (equal), data curation (equal), conceptualization (supporting), methodology (supporting), writing - review and editing (supporting);
MS: funding acquisition (supporting), writing - review and editing (equal);
JMLM: funding acquisition (lead), conceptualization (equal), supervision, methodology (lead), investigation (equal), writing - original draft, writing - review and editing (equal)

\subsection*{Acknowledgments}
Work at Weizmann was supported by the Minerva Foundation, Munich, Germany.  GHJ was funded by the National Science Foundation (grant CHE-2430408, ``Advances in Coupled Cluster Theory''; deceased PI: John F. Stanton; current PI: Alberto Perez). The authors would like to thank Prof. Mih\'aly K\'allay (Technical University of Budapest) for kindly providing a prerelease version of MRCC, and Dr. James H. Thorpe (Argonne National Laboratory) and Peter R. Franke (QTP) for helpful discussions. JMLM thanks the Quantum Theory Project at the University of Florida for their hospitality. Computational work was carried out on the Faculty of Chemistry's high-performance computing facility CHEMFARM, which is supported in part by the Ben May Center for Chemical Theory and Computation. 

%

\subsection*{Conflict of interest}

The authors declare no potential conflicts of interest.

\bibliography{postCCSDTq,FNOstuff,reviewers}

\subsection*{Supporting information}

Microsoft Excel workbook with the relevant total and interaction energies.

Additional raw data may be obtained from the corresponding author upon reasonable request.


\vspace*{12pt} 
\subsection*{Table of Contents Graphic (at magnification 100\%)}
            \begin{figure}[h!]
            \centering

\includegraphics[width=9cm]{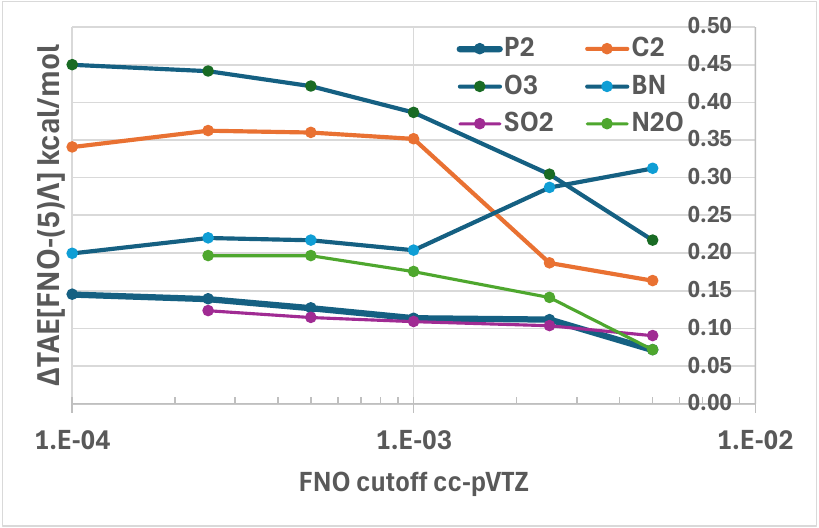}

    \addtocounter{figure}{-1}
\end{figure}


\end{document}